\documentclass{article}

\PassOptionsToPackage{numbers, sort&compress}{natbib}

\usepackage[preprint]{neurips_2022}

\usepackage{graphicx}
\usepackage{amsmath}
\usepackage{amssymb}
\usepackage{booktabs}

\usepackage[pagebackref,breaklinks,colorlinks]{hyperref}

\usepackage[capitalize]{cleveref}
\crefname{section}{Sec.}{Secs.}
\Crefname{section}{Section}{Sections}
\Crefname{table}{Table}{Tables}
\crefname{table}{Tab.}{Tabs.}

\begin{document}

\title{A Multi-Institutional Open-Source Benchmark Dataset for Breast Cancer Clinical Decision Support using Synthetic Correlated Diffusion Imaging Data}

\author{%
  Chi-en Amy Tai
  \qquad Hayden Gunraj
  \qquad Alexander Wong \\
  Vision and Image Processing Lab, University of Waterloo\\
  {\tt\small \{amy.tai, hayden.gunraj, alexander.wong\}@uwaterloo.ca}
}

\maketitle

\begin{abstract}
  Recently, a new form of magnetic resonance imaging (MRI) called synthetic correlated diffusion (CDI\textsuperscript{s}) imaging was introduced and showed considerable promise for clinical decision support for cancers such as prostate cancer when compared to current gold-standard MRI techniques. However, the efficacy for CDI\textsuperscript{s} for other forms of cancers such as breast cancer has not been as well-explored nor have CDI\textsuperscript{s} data been previously made publicly available. Motivated to advance efforts in the development of computer-aided clinical decision support for breast cancer using CDI\textsuperscript{s}, we introduce Cancer-Net BCa, a multi-institutional open-source benchmark dataset of volumetric CDI\textsuperscript{s} imaging data of breast cancer patients. Cancer-Net BCa contains CDI\textsuperscript{s} volumetric images from a pre-treatment cohort of 253 patients across ten institutions, along with detailed annotation metadata (the lesion type, genetic subtype, longest diameter on the MRI (MRLD), the Scarff-Bloom-Richardson (SBR) grade, and the post-treatment breast cancer pathologic complete response (pCR) to neoadjuvant chemotherapy). We further examine the demographic and tumour diversity of the Cancer-Net BCa dataset to gain deeper insights into potential biases. Cancer-Net BCa is publicly available as a part of a global open-source initiative dedicated to accelerating advancement in machine learning to aid clinicians in the fight against cancer. 
\end{abstract}

\section{Introduction}
\label{sec:intro}

\begin{figure}[!h]
  \centering
   \includegraphics[width=\textwidth]{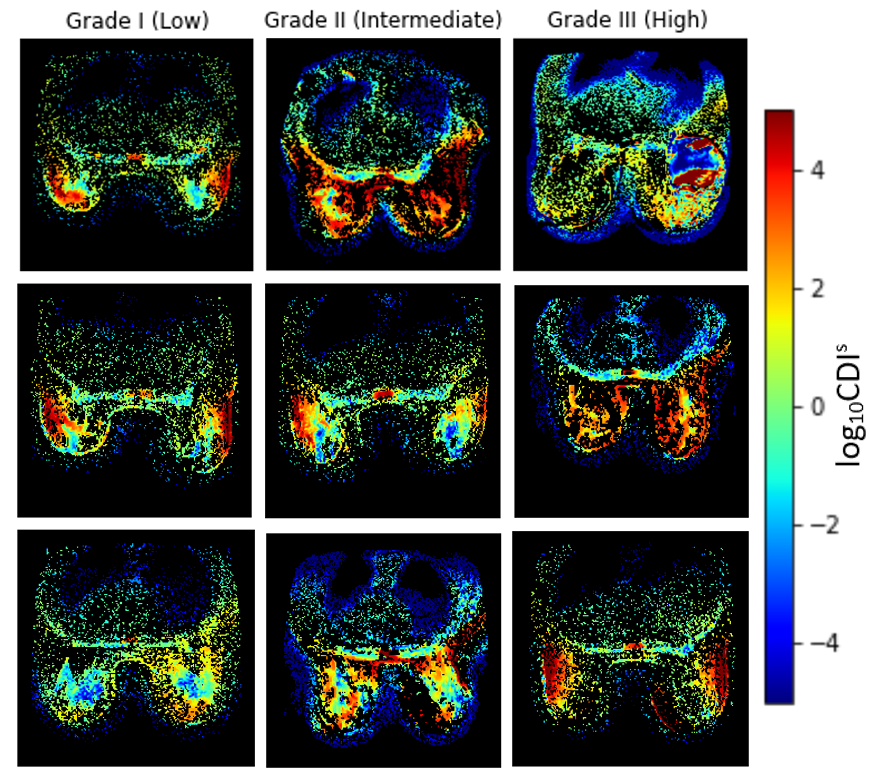}
   \caption{Example breast CDI\textsuperscript{s} images from the Cancer-Net BCa open-source benchmark dataset for each SBR grade.}
   \label{fig:grade-ex}
\end{figure}

A new form of magnetic resonance imaging (MRI) called synthetic correlated diffusion (CDI\textsuperscript{s}) imaging was recently introduced and showed considered promise for clinical decision support for cancers such as prostate cancer when compared to current gold-standard MRI techniques such as T2-weighted (T2w) imaging, diffusion-weighted imaging (DWI), and dynamic contrast-enhanced (DCE) imaging ~\cite{aa-prostate-cdis}. However, the efficacy for CDI\textsuperscript{s} for other forms of cancer such as breast cancer has not been as well-explored nor have CDI\textsuperscript{s} data been previously made publicly available. The development of computer-aided clinical decision support for breast cancer using CDI\textsuperscript{s} has begun to be analyzed and shown to have superior results compared to other gold-standard imaging for the prediction of breast cancer patient response from neoadjuvant chemotherapy prior to treatment ~\cite{ab-cancer-net-bca}. Motivated to advance efforts in the development of computer-aided clinical decision support for breast cancer using CDI\textsuperscript{s} for diagnosis, prognosis/grading, treatment planning and more, we introduce Cancer-Net BCa, a multi-institutional open-source benchmark dataset of volumetric CDI\textsuperscript{s} imaging data of breast cancer patients with detailed annotation metadata for each patient. We further examine the demographic and grade diversity of the Cancer-Net BCa dataset to gain deeper insights into potential biases. The Cancer-Net BCa benchmark dataset has been made publicly available  \footnote{https://www.kaggle.com/datasets/amytai/cancernet-bca} as a part of a global open-source initiative dedicated to accelerating advancement in machine learning to aid clinicians in the fight against cancer.

\section{Methodology}
\begin{figure}[!t]
    \tiny
    \centering
    \includegraphics[width=\textwidth]{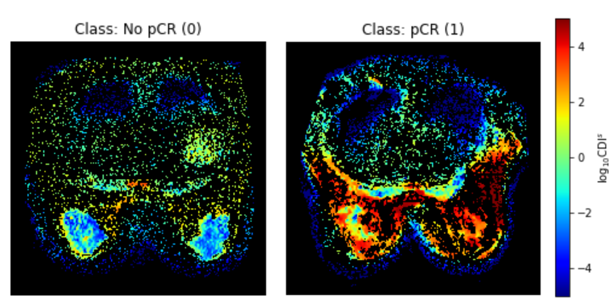}
    \caption{Example breast CDI\textsuperscript{s} images with and without pCR.}
    \label{fig:pCR-ex}
\end{figure}

To construct the Cancer-Net BCa benchmark dataset, we produced CDI\textsuperscript{s} acquisitions for a pre-treatment (T0) patient cohort of 253 patient cases across 10 institutions via the American College of Radiology Imaging Network (ACRIN) 6698/I-SPY2 study~\cite{acrin6698-data-1, acrin6698-data-2, acrin6698-data-3, acrin6698-data-4}.  More specifically, acquisitions were conducted with a four b-value imaging protocol (0 s/mm$^{2}$, 100 s/mm$^{2}$, 600 s/mm$^{2}$, 800 s/mm$^{2}$, 3-direction) on a 1.5 or 3.0 Tesla scanner using a dedicated breast radiofrequency coil. The pixel spacing for the acquisitions ranged from 0.83 mm to 2.08 mm with a median of 1.29 mm, with both slice thickness and spacing between slices ranged from 4.0 to 5.0 mm with a median of 4.0. The native and synthetic signals produced via a signal synthesizer were mixed together to obtain a final CDI\textsuperscript{s} signal~\cite{aa-prostate-cdis}. Each patient case is also associated with one of three possible SBR grades: I (Low), II (Intermediate), and III (High). Example images from each SBR type is shown in \cref{fig:grade-ex}. The pCR state after neoadjuvant chemotherapy (No pCR/pCR) is also provided for each patient, with an example of each pCR state shown in \cref{fig:pCR-ex}. 

    \begin{table}[!t]
      \centering
      \begin{tabular}{@{}lc@{}}
        \toprule
        Race & Percentage \\
        \midrule
        White & 70.8\% \\
        Black & 10.7\%\\
        Asian & 6.3\% \\
        Unknown & 11.1\% \\
        Multiple Races & 0.4\% \\      
        Native Hawaiian or other Pacific Islander & 0.4\% \\  
        American Indian or Alaska Native & 0.4\% \\  
        \bottomrule
      \end{tabular}
      \caption{Summary of race demographic in the dataset.}
      \label{tab:demographics}
    \end{table}    

\section{Results and Discussion}
\begin{figure}[!t]
      \centering
       \includegraphics[width=\linewidth]{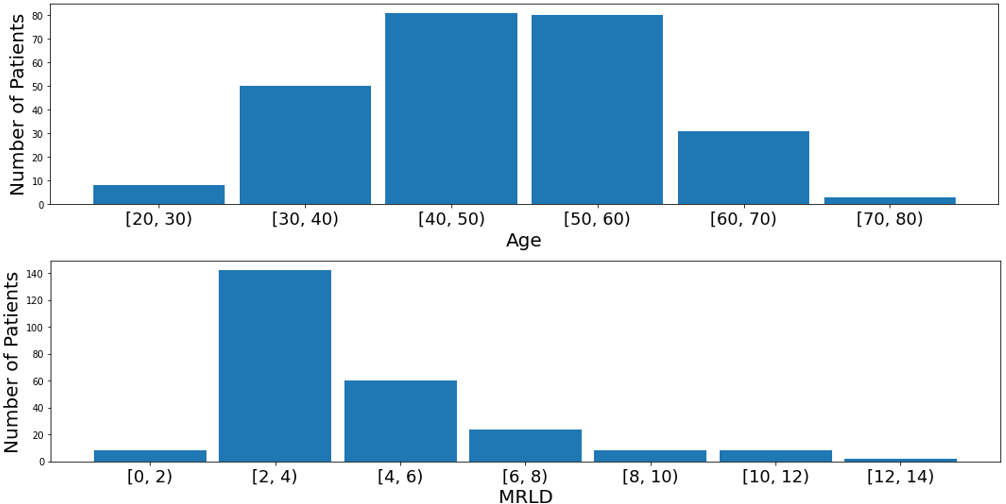}
       \caption{Distribution of the age (top) and longest diameter on the MRI (MRLD) in cm (bottom) for patients in the dataset.}
       \label{fig:age-mrld-dist}
    \end{figure}

\begin{figure}[!t]
      \centering
       \includegraphics[width=\linewidth]{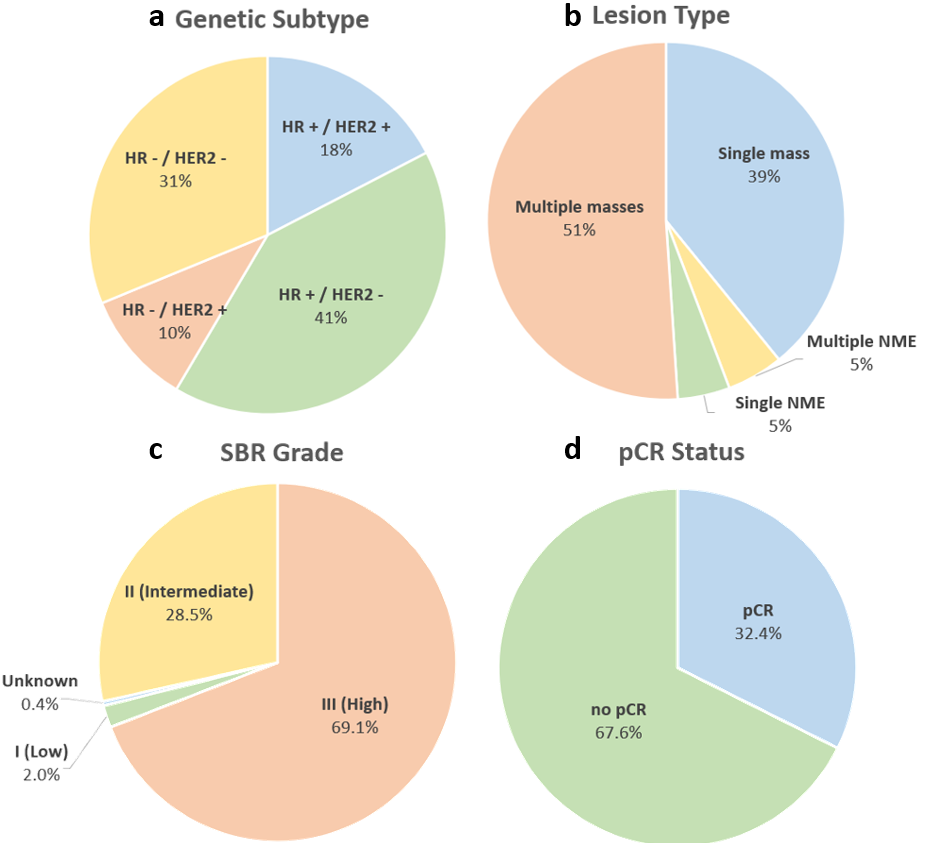}
       \caption{Patient distribution of genetic subtype (a), lesion type (b), SBR grade (c) and pCR status (d) in the dataset.}
       \label{fig:subtype-lesion-sbr-pcr}
    \end{figure}

   The demographics of the Cancer-Net BCa dataset is shown in  Table~\ref{tab:demographics}. It can be seen that the White race dominates the data, comprising of 70.8\% of the patients in the dataset, illustrating a severe race bias towards White patients. Additionally, \cref{fig:age-mrld-dist} (top), it can be seen that the majority of the patients are between 30 to 70 years old (95.7\%), indicating that very young patients ($~\leq$ 29) and very old patients ($~\geq$ 70) could be underrepresented in the dataset. On the other hand, the genetic subtype in the dataset is more fairly distributed with each subtype represented in at least 10\% of the patients whereas the lesion type is more biased towards multiple masses and single mass as seen in \cref{fig:subtype-lesion-sbr-pcr} upper left and right respectively. In addition, the longest diameter on the MRI (MRLD) is also biased towards the range of 2 to 4 cm with less representation from patients in the other diameter ranges as seen in \cref{fig:age-mrld-dist} (bottom).

The grade distribution and pCR division are shown in bottom half of Fig.~\ref{fig:subtype-lesion-sbr-pcr}, indicating an uneven distribution in SBR grade, significantly skewed towards Grade III (High) and shows that more patients with no pCR (67.6\%) compared to those who achieved pCR after neoadjuvant chemotherapy (32.4\%). Noting the demographic, grade, and pCR imbalances, it is recommended to use algorithms and strategies that account for the imbalanced dataset such as data sampling, re-balancing of the classes, and balanced loss functions. Furthermore, these imbalances should be considered when evaluating systems developed on this dataset such as with balanced metrics such as per-class precision and recall.

\bibliographystyle{unsrtnat}
{
\small

\bibliography{egbib}
}

\end{document}